\documentclass[superscriptaddress,secnumarabic,amssymb,
amsmath,nobibnotes,aps,prd,showpacs,nofootinbib]{revtex4}
\usepackage{caption}
\usepackage{subcaption}
\usepackage{graphicx}
\usepackage{bm}
\usepackage{amsmath}
\usepackage{amsfonts}
\usepackage{amssymb}
\usepackage{epstopdf}
\usepackage{color}%
\providecommand{\U}[1]{\protect\rule{.1in}{.1in}}
\newcommand{\be}{\begin{equation}}
\newcommand{\ee}{\end{equation}}

\newcommand{\mincir}{\raise
-3.truept\hbox{\rlap{\hbox{$\sim$}}\raise4.truept\hbox{$<$}\ }}
\newcommand{\magcir}{\raise
-3.truept\hbox{\rlap{\hbox{$\sim$}}\raise4.truept\hbox{$>$}\ }}

\begin{document}

 \title{Limiting curvature mimetic gravity and its relation to Loop Quantum Cosmology}

\author{Jaume de Haro\footnote{E-mail: jaime.haro@upc.edu}}
\affiliation{Departament de Matem\`atiques, Universitat Polit\`ecnica de Catalunya, Diagonal 647, 08028 Barcelona, Spain}

\author{Llibert Arest\'e Sal\'o\footnote{E-mail: Llibert.Salo@campus.lmu.de} }
\affiliation{Departament de Matem\`atiques, Universitat Polit\`ecnica de Catalunya, Diagonal 647, 08028 Barcelona, Spain}
\affiliation{Fakult{\"a}t f{\"u}r Physik, Ludwig-Maximilians-Universit{\"a}t, Theresienstr. 37, 80333 M{\"u}nchen, Germany}

\author{Supriya Pan\footnote{E-mail: span@research.jdvu.ac.in}}
\affiliation{Department of Mathematics, Raiganj Surendranath Mahavidyalaya, Raiganj, Uttar Dinajpur, West Bengal 733134, India}

\begin{abstract}
Considering as usual that the underlying geometry of our universe is well described by the spatially flat Friedmann-Lema\^{i}tre-Robertson-Walker line element, 
we 
review how  the background of  holonomy corrected Loop Quantum Cosmology (LQC)  could be obtained as a simple modified version of the mimetic gravity. We also analyze the scalar and tensor perturbations of this modified mimetic model, from which we find that at the level of tensor perturbations it is indistinguishable from General Relativity while at the  level 
of scalar perturbations, the modified mimetic model, which has the same background as LQC, does not exhibit the same properties as LQC  in the so-called {\it  deformed algebra approach.}
\end{abstract}

\vspace{0.5cm}


\maketitle

\section{Introduction}

The understanding of our universe in its extremely early phase ($\sim$ Planck scale  $\sim$ high energy regime) is not absolutely complete in the framework of Einstein's General theory of Relativity. 
Within this framework, as long as the matter sector does now allow any exotic nature, the singularity is inevitable \cite{Penrose:1964wq, Hawking1965, Hawking:1966vg, Hawking:1969sw}. A singularity $-$ the breakdown of the space-time structure $-$ means the failure of the physical laws $-$ the fundamental flaw in the description of our universe. Aside from the singularity problem, a series of early 
physics issues in the names of 
flatness problem,
horizon problem, baryon asymmetry, magnetic monopole problem 
and several others did not find any proper justification in the context 
of standard General Relativity. 
To explain such puzzles, 
it was found that the universe should undergo through a phase of
rapid accelerated expansion something like an exponential type, known as the  
inflationary paradigm \cite{guth, Linde:1981mu} and, consequently, 
the theory of inflation became quite successful in explaining most 
of the serious issues. Nonetheless the singularity could not be 
framed within such formalism and stayed as a shortcoming of the 
General theory of Relativity. This drived the scientific 
community towards the quantum theory of gravity, since     
dealing with the evolution of the universe 
in the Planck's scale quantum effects should be extremely important
and effective in comparison to our classical description of the 
General Relativity. The investigations in the last several years  
deal with two promising theories of quantum gravity, namely \textit{String theory} \cite{Polchinski}
and \textit{Loop Quantum Gravity} (LQG) \cite{Rovelli:1997yv}. In the current work we will consider the cosmological descriptions of the later one, that means, \textit{Loop Quantum Cosmology} (LQC), see \cite{as} for a review on the developments of LQC. 

\

{{} It has been found that the matter-ekpyrotic scenario in  LQC \cite{cw,haa} in the so-called {\it deformed algebra approach} \cite{grain},  which contains a phase transition in the contraction regime and as a consequence allows our universe to be reheated via the particle production mechanism, depicts a viable cosmology whose theoretical values of the power spectrum, spectral index and its running  agree with their  observational estimations extracted from the joint analysis of BICEP2/Keck Array and Planck \cite{planck}. We also refer to some earlier works where the authors show that inflation \cite{Date:2004yz} and bounce \cite{Date:2004fj} are generically obtained in this context. }

\

A suitable connection between LQC and the modified gravity theories was established in a series  of articles \cite{helling,ds09,haro12, llkw} showing that, 
in the framework of the usual spatially flat Friedmann-Lema\^{i}tre-Robertson-Walker (FLRW) universe, the holonomy corrected LQC model can be equivalent to modified gravity if one works with an invariant scalar that depends only on the square of the Hubble parameter  \cite{helling,ds09,haro12, llkw}.  One may recall that such a scalar is already known in some viable modified gravity models, for instance, the scalar could be the torsion that appears in the teleparallel gravity models where the spacetime is equipped with an unusual  Weitzenb\"ock connection \cite{w}, and a preferred orthonormal basis in the tangent bundle of the spacetime manifold must be selected. As a second thought, we could recall 
the extrinsic curvature scalar in the context of 
the Arnowitt-Deser-Misner (ADM) formalism of GR \cite{ADM}. One may find that, in the  above two frameworks, if the background universe is perfectly described by the spatially flat FLRW spacetime, for the synchronous co-moving coordinates \cite{ha17} the scalar becomes `$-6H^2$'. {Moreover, using the extrinsic curvature scalar, it has been showed in \cite{ha17} that one obtains the same equation of scalar perturbations as the one obtained in holonomy corrected LQC using the
{\it deformed algebra approach} (see for instance \cite{grain}), which does not happen using the torsion provided by the Weitzenb\"ock connection \cite{haro}.}

\

Although both the modified gravity models mentioned above are believed to be the viable alternatives for describing the expansive history of the universe, one may recall that both of them share a common trouble $-$ the  preferred coordinate system.  Locally the teleparallel gravity is not Lorentz invariant \cite{lsb} and, concerning the extrinsic curvature scalar in the ADM formalism of GR, it requires to fix the slicing \cite{ha17}. Thus, 
both the modified gravity theories have essentially some shortcomings. It might be important to refer to some recent 
investigations which argue that if the spin connection is also considered along with the pure-tetrad formalism for the teleparallel gravity the local invariance problem might be resolved \cite{Krssak:2015oua, Hohmann:2018rwf}. However, the consequences of such proposals, where in particular the spin connection is important for the teleparallel models, need some more time for a decisive conclusion towards this direction.  
On the other hand, for the ADM formalism of GR we are not concerned with any kind of proposal, at least at the time of writing this paper. Thus, one may readily conclude that the scalar we find in both these modified gravity theories is not gauge invariant. Certainly the gauge invariant scalars are of utmost important. 
However, the construction of such gauge invariant scalars is not so easy, nevertheless they are not impossible. 
These scalars can be built with the  Carminati-McLenaghan invariants \cite{hp18} (see \cite{cm} for the definition of these invariants)  or the d'Alembertian of a mimetic field \cite{mukhanov,mukhanov1}.

\

Once such gauge invariant scalars are found, the realization of the cosmic bounce becomes simple by working in the phase space $(H,\rho)$, where $H$ denotes de Hubble parameter and $\rho$ the energy density of the FLRW universe.
 In particular, the simplest  route to build  such bouncing backgrounds is to consider the closed curves in the $(H,\rho)$ phase space crossing  the $H=0$ axis, at least twice. 
Now, for a  given $f$-theory, the modified Friedmann equation becomes a first
order differential on $f$, that relates $f$ and $\rho$, both  as a function of the invariant scalar. The solution of the modified Friedmann equation returns the corresponding $f$ theory that leads to the background depicted by the corresponding curve. We remark that the simplest closed curve in the  $(H,\rho)$ phase space is an ellipse which depicts the holonomy corrected Friedmann equation in LQC (see for details \cite{hp18}). 

\

In the present work we use the scalar provided by the modified mimetic gravity
(the D'Alembertian of the mimetic field with a negative sign)
and show, as already done in \cite{llkw,mukhanov1,Norbert},  that there is an $f$-theory which reproduces exactly the same background as the holonomy corrected LQC. We extend our analysis considering both the scalar and tensor perturbations. 
{{} Since in modified mimetic gravity 
 the classical hypersurface deformation algebra is obtained (see for instance  \cite{Norbert}), as a consequence, the equations for perturbations coming from modified mimetic gravity might not be the same as in the deformed algebra approach of LQC (see for instance \cite{bojowald1,bojowald2} for a review of the ``state-of-the-art" of  the problem of covariance in canonical quantum gravity)}. 
In fact, 
our results show that at the level of scalar perturbations 
the modified mimetic model differs with LQC because it contains extra complicated terms. And, for the tensor perturbations, as has recently been pointed out in \cite{hnk17}, the mimetic field has no influence and, thus, one obtains the well-known tensor perturbed equations for General Relativity. 

\

The paper has been organized in the following way. In section \ref{sec-2} we introduce the modified mimetic model and its evolution in the background level. The perturbation equations have been presented in section \ref{sec-perturbed}, where in particular the scalar and tensor perturbations are shown in subsections \ref{sec-scalar-perturbed} and
\ref{sec-tensor-perturbed} respectively. Finally, we close the entire work in section \ref{sec-discuss} with the main findings of the work.   
The units used throughout the paper are $\hbar=c=1=M_{pl}=1$, where $M_{pl}$ is the reduced Planck's mass (the other notations have their usual meanings), with the convention that a temporal vector $v_{\mu}$ satisfies $v_{\mu}v^{\mu}<0$. The meanings of some frequently used quantities in this work are, (a) $\phi_{,\mu}\equiv\partial_{\mu}\phi=\nabla_{\mu}\phi~$, (b) $\bar{\phi}$ as the unperturbed part of $\phi~$; (c) $f_{\chi}$ is the derivative of $f$ with respect to $\chi$ and $V_{\phi}$ is the derivative of $V$ with respect to $\phi~$.

\section{Modified Mimetic Gravity}
\label{sec-2}

In recent time, cosmology with mimetic gravity is getting an impressive 
attention \cite{Chamseddine:2013kea, Golovnev:2013jxa, Nojiri:2014zqa, Leon:2014yua, Odintsov:2015wwp, Casalino:2018tcd, Brahma:2018dwx} and is in progress both in the theoretical and observational level (also see a recent review in this direction \cite{Sebastiani:2016ras}). Here we allow an extension of the original 
mimetic gravity in terms of  the following action
\begin{eqnarray}\label{action}
 S=\int\sqrt{-g}\left(\frac{1}{2}R+\lambda(\varphi_{,\mu}\varphi^{,\mu}+1)+f(\chi)    +{\mathcal L}_{matt}\right) d^4x,
 \end{eqnarray}
where  $R$ is the scalar curvature, $\varphi$ is the mimetic field satisfying $\varphi^{,\mu}\varphi_{,\mu} \equiv \nabla^{\mu}\varphi\nabla_{\mu}\varphi
 =-1$;  $\chi\equiv -\Box \varphi=-\nabla^{\mu}\varphi_{,\mu}~$; $\lambda$ is a Lagrange multiplier. Further, we have assumed that the matter sector of the universe is filled with a scalar field $\phi$ having potential $V(\phi)$, which is minimally coupled 
 to gravity whose Lagrangian is given by
 \begin{eqnarray}
  {\mathcal L}_{matt}= -\frac{\phi_{,\mu}\phi^{,\mu}}{2}-V(\phi).
  \end{eqnarray}
Although one can extend the action (\ref{action}) for $F(R)$ gravity \cite{noo} 
with the replacement $R \rightarrow F (R)$, in this work we only consider the simplest case  described in the above action (\ref{action}). 
The dynamical equations making the variation of (\ref{action}) with respect to $g_{\mu\nu}$ are \citep{mukhanov1}
\begin{eqnarray}\label{dyn}
G_{\mu\nu}=T_{\mu\nu}-\tilde{T}_{\mu\nu}-2\lambda \varphi_{,\mu}\varphi_{,\nu},
\end{eqnarray}
where $G_{\mu\nu}=R_{\mu\nu}-\frac{1}{2}g_{\mu\nu}R$ is the Einstein tensor, 
$T_{\mu\nu}=\phi_{,\mu}\phi_{,\nu}-\left(\frac{1}{2}\phi_{,\alpha}\phi^{,\alpha}+V(\phi)   \right)g_{\mu\nu}$ is the stress tensor and 
\begin{eqnarray}
\tilde{T}_{\mu\nu}\equiv g_{\mu\nu} \bigl(\chi f_{\chi}-f-\varphi_{,\alpha}\chi^{,\alpha}f_{\chi\chi} \bigr) +f_{\chi\chi} \bigl(\varphi_{,\nu}\chi_{,\mu} +\varphi_{,\mu}\chi_{,\nu} \bigr).
\end{eqnarray}

The dynamical equation for the mimetic field  is obtained after performing the variation of the action with respect to $\varphi$ as follows 
\begin{eqnarray}\label{const}
\partial_{\mu}\Bigl[\sqrt{-g}\left(f_{\chi\chi}\chi^{,\mu}+2\lambda\varphi^{,\mu}    \right)   \Bigr]=0,
\end{eqnarray}
and the variation of the scalar field $\phi$ leads to the well-known conservation equation
\begin{eqnarray}\label{conser}
-\Box \phi+ V_{\phi}=0.
\end{eqnarray}

Note that the dynamical equation for the scalar field (\ref{conser}) could also be deduced   from the conservation equation $\nabla_{\mu}T^{\mu}_{\nu}=0$, and
the one for the mimetic field must also be obtained by taking the divergence of (\ref{dyn}). From the Bianchi identity $\nabla_{\mu}G^{\mu}_{\nu}=0$ and
the conservation equation $\nabla_{\mu}T^{\mu}_{\nu}=0$, it is verified that $\nabla_{\mu}(\tilde{T}^{\mu}_{\nu}+2\lambda \varphi^{,\mu}\varphi_{,\nu})=0$. Further, following some algebraic calculations, one can see that the equation 
$\varphi^{\nu}\nabla_{\mu}(\tilde{T}^{\mu}_{\nu}+2\lambda \varphi^{,\mu}\varphi_{,\nu})=0$ is equivalent to equation (\ref{const}).

\subsection{Background equations}
\label{sec-background}

In this section we explicitly describe the background equations for the action  (\ref{action}) in a spatially flat Friedmann-Lema{\^\i}tre-Robertson-Walker (FLRW) geometry. We work with the synchronous observers, that means, the line element takes the form $ds^2=-dt^2+a^2\delta_{ij}dx^idx^j$, where $a(t)$ is the expansion scale factor 
of the universe. The simplest solution of
$\bar\varphi_{,\mu}\bar\varphi^{,\mu}=-1$, is $\bar\varphi(t)=t$, leading to $\bar\chi=3H$. Thus,  equation (\ref{const}) becomes
\begin{eqnarray}\label{varphi}
\partial_t \left[a^3(3\dot{H}\bar f_{\chi\chi}+2\bar\lambda) \right]=0,
\end{eqnarray}
whose general solution is {{} $\bar\lambda=\frac{C}{2} a^{-3}-\frac{3}{2}\dot{H}\bar f_{\chi\chi}$.} For simplicity, we take $C=0$, thus, the $0-0$ component of the field equations (\ref{dyn}) becomes
\begin{eqnarray}\label{modfriedman}
\rho=3H^2+\bar f -3H\bar f_{\chi},
\end{eqnarray}
which depicts a curve in the phase space of $(H,\rho)$.
The $i-0$ component identically vanishes whereas the $i-i$ component yields the Raychaudhuri equation
\begin{eqnarray}\label{raychauduri}
\left(1-  \frac{3}{2}\bar f_{\chi\chi} \right)\dot{H}=-\frac{1}{2}(\rho+P),
\end{eqnarray}
which is equivalent to the conservation equation $\dot{\rho}=-3H(\rho+P)$, where $P$ is the pressure associated with the energy density $\rho$.

Therefore, for given a curve $\rho=\bar g(3H)=\bar g(\bar\chi)$, in order to obtain the corresponding $\bar f(\bar\chi)$ one has to solve the first order differential equation (\ref{modfriedman}) written as 
\begin{eqnarray}
\bar\chi \bar f_{\chi}-\bar f-\frac{1}{3}\bar\chi^2+\bar g(\bar\chi)=0,
\end{eqnarray}
whose solution is
\begin{eqnarray}
\bar f(\bar\chi)=+\frac{1}{3}\bar\chi^2-\bar\chi\int\frac{\bar g(\bar\chi)}{\bar\chi^2}d\bar\chi.
\end{eqnarray}

As an example, we consider the holonomy corrected Friedmann equation in LQC, i.e.,  we take the following ellipse 
\begin{eqnarray}\label{flqc}
\rho=\bar g(\bar\chi)=\frac{\rho_c}{2}\left(1\pm\sqrt{1-\frac{4\bar\chi^2}{3\rho_c}}   \right)\Longleftrightarrow H^2=\frac{\rho}{3}\left(1-\frac{\rho}{\rho_c}\right),
\end{eqnarray}
which leads to the mimetic $f$-theory 
\begin{eqnarray}\label{fLQC}
\bar f(\bar \chi)=\frac{1}{3}\bar\chi^2+\frac{\rho_c}{2}\left(1-\sqrt{1-s^2}-s\arcsin(s)  \right),
\end{eqnarray}
first proposed by \cite{mukhanov1}, where $s\equiv \frac{2}{\sqrt{3\rho_c}}\bar\chi$ and the functions $\sqrt{1-s^2}$,  $\arcsin(s)$ are bi-valued  \cite{hp18}. In order to be well defined we have to take the following  prescription. The ellipse has two branches where the upper part corresponds to 
$\rho=\frac{\rho_c}{2}\left(1+\sqrt{1-\frac{4\bar\chi^2}{3\rho_c}} \right)$ and the lower one is depicted by 
$\rho=\frac{\rho_c}{2}\left(1-\sqrt{1-\frac{4\bar\chi^2}{3\rho_c}} \right)$. We choose
the sign of the square root as positive (respectively negative) in the lower (respectively upper) branch and 
$ \arcsin(s)\equiv \int_0^s \frac{1}{\sqrt{1-{\bar s}^2}}  {d\bar{s}}$ in the lower branch, while $\arcsin(s)\equiv \int_0^s \frac{1}{\sqrt{1-{\bar s}^2}}  {d\bar{s}}+\pi$ in the upper one, having the same criteria for the sign of the square root, thus obtaining that the function $\bar f$ is continuous throughout the ellipse.

Finally, a simple calculation shows that for this particular theory the $0-0$ and $i-i$ equations respectively become
\begin{eqnarray}
H^2=\frac{\rho}{3}\left(1-\frac{\rho}{\rho_c}\right),
\end{eqnarray}
and
\begin{eqnarray}
\dot{H}=-\frac{\rho+P}{2}\left(1-\frac{2\rho}{\rho_c}\right),
\end{eqnarray}
which are nothing but the modified field equations in the 
holonomy corrected LQC \cite{as}. As we will see later our theory differs to some extent from Loop Quantum Cosmology. Hence, we can not include the name of LQC in the $f$-theory given by \eqref{fLQC} and we will call it simply {\it Limiting curvature mimetic theory}, in analogy to \cite{mukhanov1}.

\

A final remark is in order:
The simplest route to obtain the background equations goes as follows \cite{llkw}. Let us consider the metric $ds^2=-N^2dt^2+a^2\delta_{ij}dx^idx^j$, where we introduce the lapse function $N$. In this case the constraint $\varphi^{,\mu}\varphi_{,\mu}=-1$ reads as $\dot{\bar\varphi}=N$ and, thus,
$\bar\chi=\frac{3H}{N}$. Then, introducing these results in (\ref{action}) we obtain the reduced action
\begin{eqnarray}\label{action-reduced}
S_{red}=\int a^3N\left(  -\frac{3H^2}{N^2}+ \bar{f}\left(\frac{3H}{N}  \right)+{\mathcal L}_{matt}   \right) dt,
\end{eqnarray}
where the matter Lagrangian is
\begin{eqnarray}
{\mathcal L}_{matt} =\frac{{\dot{\bar{\phi}}}^2}{2N^2}-V(\bar\phi).
\end{eqnarray}

We note that performing the variation  of the reduced action (\ref{action-reduced}) with respect to the lapse function $N$ and at the end making $N=1$, without any loss of generality, one obtains the Friedmann equation (\ref{modfriedman}). Moreover, performing the  temporal derivative of (\ref{action-reduced}) and using the conservation equation $\dot{\rho}=-3H(P+\rho)$, one may find the Raychaudhuri equation (\ref{raychauduri}). To finish this remark it is interesting to mention the deduction of these equations made in \cite{Norbert} using the Hamiltonian formalism.

\section{Perturbations}
\label{sec-perturbed}

The behavior of any cosmological theory in the large scale of the universe is the most important  subject for investigation. Thus, following the evolutions of the mimetic modified gravity model at the  background level in section \ref{sec-background}, 
in this section we calculate the scalar and tensor perturbations for the present model using the longitudinal gauge.

\subsection{Scalar Perturbations}
\label{sec-scalar-perturbed}

{
In theory of perturbations, dealing with  scalar perturbations,  the most general line element is 
\begin{eqnarray}
d^2s=a^2(\eta)\left\{-(1+2\Phi)d\eta^2+2B_{, i}dx^id\eta +[(1-2\Psi)\delta_{ij}+2E_{,ij}]    \right\},
\end{eqnarray}
where $\eta$ is the cosmic time.
From the fuctions $\Phi$, $\Psi$, $B$ and $E$ one can construct the so-called gauge-invariant Bardeen's potentials \cite{mfb}
\begin{eqnarray}
\Phi^{gi}=\Phi+\frac{1}{a}[(B-E')a]', \qquad \Psi^{gi}=\Psi-\mathcal{H}(B-E'),
\end{eqnarray} 
where ${\mathcal H}$ is the conformal Hubble parameter. In the same way, for any scalar $q=\bar{q}+\delta q$ one could build up the gauge-invariant 
$\delta q^{gi}=\delta q+\bar{q}'(B-E')$. Then, choosing the Newtonian gauge, for which $B=E=0$, we can see that the variables $\Phi$, $\Psi$ and the perturbations
of any scalar are gauge-invariants. Using that gauge and the cosmic time, the
}
line element can be written as 
\begin{eqnarray}
ds^2=-(1+2\Phi)dt^2+(1-2\Phi)a^2\delta_{ij}dx^idx^j,
\end{eqnarray}
because in mimetic gravity $\delta(T_{i}^{j}-\tilde{T}_{i}^{j}-2\lambda \varphi_{i}\varphi^{j})$ vanishes for $i\not=j$ \cite{mukhanovbook}. One may identify $\Phi$ and 
$\Psi$ as the perturbation quantities. 
We perturb the scalar field as $\varphi=t+\delta\varphi$, then the equation $\varphi_{,\mu}\varphi^{,\mu}=-1$ leads to $\Phi=\delta\dot{\varphi}$.

At the  linear order, a simple calculation leads to  
{{}\begin{eqnarray}
\chi= 3H-3(\dot{\Phi}+H\Phi)-\frac{1}{a^2}\Delta\delta\varphi.
\end{eqnarray}}
And perturbing the equation (\ref{dyn}), the 
$i-0$, $i-i$ and $0-0$ equations respectively take the following forms:
\begin{eqnarray}
\dot{\Phi}+H\Phi=\frac{1}{2}\dot{\bar\phi}\delta\phi-\frac{1}{2}\bar f_{\chi\chi}\delta\chi,
\end{eqnarray}
\begin{eqnarray}
2\left(  \ddot{\Phi}+4H\dot{\Phi}+(3H^2+2\dot{H})\Phi
\right)=-\dot{\bar\phi}(\Phi\dot{\bar\phi}-\delta\dot{\phi})-\bar V_{\phi}\delta\phi -
\delta(\chi f_{\chi}-f -\varphi^{,\mu}\chi_{,\mu}f_{\chi\chi}),
\end{eqnarray}
\begin{eqnarray}
2\left(3{ H}^2\Phi+3{ H}\dot{\Phi}-\frac{1}{a^2}\Delta \Phi\right)=\dot{\bar\phi}(\Phi\dot{\bar\phi}-\delta\dot{\phi})-\bar V_{\phi} \delta\phi-
\delta(\chi f_{\chi}-f +\varphi^{,\mu}\chi_{,\mu}f_{\chi\chi})
+2\delta\lambda ~.
\end{eqnarray}

 On the other hand, 
 the general solution of the  equation (\ref{const}) is
 \begin{eqnarray}
 f_{\chi\chi}\chi^{,\mu}+2\lambda \varphi^{,\mu}=\xi^{\mu},
\end{eqnarray}  
 where $\xi^{\mu}$ is a vector field satisfying $\nabla_{\mu}\xi^{\mu}=0$. Then, one has 
 \begin{eqnarray}
 2\lambda =-\xi^{\mu}\varphi_{,\mu}+f_{\chi\chi}\chi^{,\mu}\varphi_{,\mu},
 \end{eqnarray}
and by perturbing it one obtains
 \begin{eqnarray}
 2\delta\lambda=\delta\xi_0-\partial_t(\bar f_{\chi\chi}\delta\chi)+\dot{\bar f}_{\chi}\delta\dot{\varphi},
 \end{eqnarray}
which leads to 
\begin{eqnarray}\label{1}
\delta\lambda=\frac{1}{2}\delta\xi_0+\frac{\bar f_{\chi\chi}}{2}(3\ddot{\Phi}+\frac{1}{a^2}\Delta \Phi)+\frac{3}{2}(H\bar f_{\chi\chi}+\dot{\bar f}_{\chi\chi})\dot{\Phi}
+\frac{3}{2}(2\dot{H}\bar f_{\chi\chi} +H\dot{\bar f}_{\chi\chi} )\Phi+(\frac{1}{2}\dot{\bar f}_{\chi\chi}-H\bar f_{\chi\chi})\frac{1}{a^2}\Delta \delta\varphi.
\end{eqnarray}

Now, adding the equations $0-0$ and $i-i$ and using $i-0$, one gets
 \begin{eqnarray}
 \ddot{\Phi}-\frac{1}{a^2}\Delta \Phi+H\dot{\Phi}+2\dot{H}\Phi=\ddot{\bar\phi}\delta\phi+\delta\lambda.
\end{eqnarray}

Then, writing $i-0$ as follows,
\begin{eqnarray}\label{2}
\ddot{\bar\phi}\delta\phi=2\frac{\ddot{\bar\phi}}{\dot{\bar\phi}}(\dot{\Phi}+H\Phi)(1-\frac{3}{2}\bar f_{\chi\chi})
-\frac{\ddot{\bar\phi}}{\dot{\bar\phi}}\bar f_{\chi\chi}\frac{1}{a^2}\Delta \delta\varphi,
\end{eqnarray}
 collecting terms and introducing the notation $\Omega\equiv \frac{1}{1-\frac{3}{2}\bar f_{\chi\chi}}$, one gets {{} the equation for the potential $\Phi$,
\begin{align}\label{Phieq}
\ddot{\Phi}-\frac{\Omega}{a^2}\Delta \Phi+\left(H-2\frac{\ddot{\bar\phi}}{\dot{\bar\phi}}-\frac{\dot{\Omega}}{\Omega}\right)\dot{\Phi}+
\left(2\left(\dot{H}-H\frac{\ddot{\bar\phi}}{\dot{\bar\phi}}\right)-H\frac{\dot{\Omega}}{\Omega}\right){\Phi}=\frac{\Omega}{2}\left[
\delta\xi_0+
{\dot{\bar{\phi}}}^2\partial_t\left( \frac{\bar{f}_{\chi\chi}\Delta\delta\varphi}{a^2{\dot{\bar{\phi}}}^2 } \right)\right],
\end{align}
where $\delta\xi_0$ satisfies the linearized equation of (\ref{const}), i.e.,
\begin{eqnarray}\label{Y}
\delta\dot{\xi}_0+3H\delta{\xi}_0 +
\frac{\bar{f}_{\chi\chi}}{a^2}\left[{3}\left(\Delta\dot{\Phi}+H\Delta\Phi  \right)+{3}\dot{H}\Delta\delta\varphi+\frac{1}{a^2}\Delta^2\delta\varphi
\right]=0.
\end{eqnarray}

Note that the equation (\ref{Phieq}) is different from the one obtained in holonomy corrected LQC using the "deformed algebra approach" \cite{grain},  due to the fact that 
 the term on the right hand side of (\ref{Phieq}) does not vanish.
 
\

Moreover, we also need the linearized conservation equation  (\ref{conser}):
 \begin{eqnarray}\label{linearization}
\delta\ddot{\phi}+3H\delta\dot{\phi}-\frac{1}{a^2}\Delta \delta\phi+V_{\phi\phi}\delta\phi-4\dot{\bar{\phi}}\dot{\Phi}+2V_{\phi}\Phi=0.
\end{eqnarray}

 \
 
 Now, once we have all the dynamical equations, we can identify the dynamical variables, which in this case are $\delta\phi$ and $\delta\varphi$, because from the constraint
 $\varphi_{,\mu}\varphi^{,\mu}=-1$ we have obtained the relation $\Phi=\delta\dot\varphi$. Then, the dynamical equations are composed by the equation $i-0$ and
 (\ref{linearization}), leading to the coupled
 system:
 \begin{eqnarray}\label{system}
 \left\{\begin{array}{ccc}
 \delta\ddot{\varphi}+H\delta\dot{\varphi}-\frac{c_s^2}{a^2}\Delta\delta\varphi &=&\frac{\Omega}{2}\dot{\bar\phi}\delta\phi\\
 \delta\ddot{\phi}+3H\delta\dot{\phi}-\frac{1}{a^2}\Delta \delta\phi+(V_{\phi\phi}-2{\dot{\bar\phi}}^2{{}\Omega})\delta\phi &=& 
 \frac{4{{}\dot{\bar\phi}}c_s^2}{a^2}\Delta\delta\varphi
 -2(2{{}H\dot{\bar\phi}}+V_{\phi})\delta\dot{\varphi},
 \end{array}\right.
 \end{eqnarray}
 where the square of the velocity of sound is given, as in \cite{fgm17}, by 
 \begin{eqnarray} \label{cs2}
 c_s^2=\frac{\Omega}{2}\bar{f}_{\chi\chi}=\frac{\frac{1}{2}\bar{f}_{\chi\chi}}{1-\frac{3}{2}\bar{f}_{\chi\chi}},
 \end{eqnarray}
 which  exhibits the well-known gradient instability of the mimetic field \cite{hnk17, fgm17}.
 Moreover, since for the $f$-theory which leads to the same background of holonomy corrected LQC (eq. (13)), one has
 $\bar{f}_{\chi\chi}=\frac{2}{3}\left(1-\frac{1}{\Omega}  \right)=-\frac{4\rho}{3(\rho_c-2\rho)}$ one can see that, for this theory,  $c_s^2$ is always negative except
 for $\rho=0$. So, this shows a clear difference from this theory in LQC, since LQC does not have a gradient instability at all times, but only for the upper half of the ellipse ($\frac{\rho_c}{2}<\rho<\rho_c$)\cite{haa}.

 Then, working in the Fourier space, given a background $\dot{\bar\phi}$ choosing initial conditions at a given time $t_0$, $\delta\phi_k(t_0)$, 
 $\delta\dot \phi_k(t_0)$, $\delta\varphi_k(t_0)$ and  $\delta\dot \varphi_k(t_0)$, solving the system (\ref{system}) one obtains the evolution of the two degrees of freedom
 $\delta\phi_k$ and $\delta\varphi_k$.

 
 }


 \

 To end this section we will calculate the Mukhanov-Sasaki (M-S) equation for scalar perturbations. First of all, note that  the equation $i-0$ could be written as
{{}\begin{eqnarray}
\frac{d}{dt}\left(\frac{a\delta\dot\varphi}{H}  \right)=\frac{a\Omega{\dot{\bar{\phi}}}^2}{2H^2}
\left[\frac{H\delta\phi}{\dot{\bar{\phi}}}+\delta\dot\varphi+\frac{H}{a^2{\dot{\bar{\phi}}}^2}\bar{f}_{\chi\chi}\Delta\delta\varphi  \right].
\end{eqnarray}

 On the other hand, the equation $0-0$ has the form
 \begin{eqnarray}
2\left(3{ H}^2\delta\dot\varphi+3{ H}\delta\ddot{\varphi}-\frac{1}{a^2}\Delta \delta\dot\varphi\right)=\dot{\bar\phi}(\delta\dot\varphi\dot{\bar\phi}-\delta\dot{\phi})-\bar V_{\phi} \delta\phi-
3H\bar{f}_{\chi\chi}\delta\chi
+\delta\xi_0,
\end{eqnarray}}
and
after a cumbersome calculation
one can see that it is equivalent to the following one
{{}
\begin{eqnarray}
\frac{1}{a^2}\Delta \delta\dot\varphi=\frac{{\dot{\bar{\phi}}}^2}{2H}\frac{d}{dt}\left[\frac{H\delta\phi}{\dot{\bar{\phi}}}+\delta\dot\varphi+
\frac{H}{a^2{\dot{\bar{\phi}}}^2}\bar{f}_{\chi\chi}\Delta\delta\varphi  \right]-\frac{1}{2}\delta\xi_0
-\frac{\dot{\bar{\phi}}^2}{2}\partial_t\left( \frac{\bar{f}_{\chi\chi}\Delta\delta\varphi}{a^2{\dot{\bar{\phi}}}^2 } \right).
\end{eqnarray}
}

{ Now, introducing the gauge-invariant (recall that we are working in the Newtonian gauge) variables,}
{{}\begin{eqnarray}
v=a\left(\delta\phi+\frac{\dot{\bar{\phi}}}{H}\delta\dot\varphi +  \frac{1}{a^2{\dot{\bar{\phi}}}}\bar{f}_{\chi\chi}\Delta\delta\varphi  \right), \quad z=a\frac{\dot{\bar{\phi}}}{H}, \quad
u=\frac{2\delta\dot\varphi}{\dot{\bar{\phi}}\sqrt{\Omega}}, \quad \mbox{and} \quad \theta=\frac{1}{z\sqrt{\Omega}},
\end{eqnarray} }
and using the conformal time one obtains the M-S equations
 {{}\begin{eqnarray}
 \sqrt{\Omega}\Delta u=z\left(\frac{v}{z}\right)'-\frac{a^3}{\bar{\phi}'}\delta\xi_0
 -{{\bar{\phi}}'}\left( \frac{\bar{f}_{\chi\chi}\Delta\delta\varphi}{(\bar{\phi}')^2 } \right)'
 ,\qquad \theta\left(\frac{u}{\theta}\right)'=\sqrt{\Omega} v.
 \end{eqnarray}
 }
 
 Performing the Laplacian in the second equation and using the first one, we obtain
 {{}\begin{eqnarray}
 v''-\Omega\Delta v-v\frac{z''}{z}={{}\frac{1}{z}}\left[\left(\frac{za^3}{\bar{\phi}'}\delta\xi_0\right)'+
\left(z{{\bar{\phi}}'}\left( \frac{\bar{f}_{\chi\chi}\Delta\delta\varphi}{(\bar{\phi}')^2 } \right)'\right)'\right] ,
 \end{eqnarray}
 and inserting the second equation in the first one we get the equivalent to the equation (\ref{Phieq})
 \begin{eqnarray}
 u''-\Omega\Delta u-u\frac{\theta''}{\theta}=\sqrt{\Omega}\frac{a^3}{\bar{\phi}'}\delta\xi_0+
\sqrt{\Omega}{{\bar{\phi}}'}\left( \frac{\bar{f}_{\chi\chi}\Delta\delta\varphi}{(\bar{\phi}')^2 } \right)',
 \end{eqnarray}
 which do not coincide with the M-S in LQC for scalar perturbations \cite{ha14a} because the right hand side of these equations does not vanish.}

\

{

We note that the equations deduced in this last section have been written such that the left-hand side coincides with the ones of the LQC in the deformed algebra approach, but the right-hand side differs from zero, which shows that both approaches are different. For this reason, the velocity of sound for the different variables in general cannot be simply extracted from these equations. However, the coupled system \eqref{system} is probably the most direct way to interpret the sound speeds, obtaining namely $1$ for the scalar field and, as in \cite{fgm17}, \eqref{cs2} for the mimetic field.

}

\subsection{Tensor Perturbations}
\label{sec-tensor-perturbed}

Concerning the tensor perturbations for the modified mimetic model, 
the perturbed metric is given by 
\begin{eqnarray}
ds^2=-dt^2+a^2(\delta_{ij}-h_{ij})dx^idx^j,
\end{eqnarray}
where $h_{ij}$ is a symmetric,  traceless and transverse tensor, that means, $h_i^i=\partial_ih^{ij}=0$. The constraint in this case now becomes $\varphi^{,\mu}\varphi_{,\mu}=-1$ which  leads to $\delta\dot\varphi=0$ and, thus, at linear order one arrives at $\chi=3H-\frac{1}{a^2}\Delta \delta\varphi$. 
 
From this result, we can see that for the tensor perturbations the $i-0$ equation  becomes $\Delta\delta\varphi=0$, which leads to $\delta\chi=0$.  As a consequence, the right hand side of the $i-j$ equation vanishes leading to $\delta G_i^i=0$, which is equivalent to the
well-known equation of tensor perturbations in GR,
\begin{eqnarray}
\ddot{h}_{i}^{j}+3H\dot{h}_{i}^{j}-\frac{1}{a^2}\Delta h_{i}^{j}=0,
\end{eqnarray}
as also shown in \cite{hnk17} for mimetic gravity. This is a feature of this theory which is indeed different from the  Teleparallel LQC \cite{cai,ha14a}, Extrinsic curvature LQC \cite{ha17}
or even from the holonomy corrected LQC  \cite{caitelleau},
in all of which the equation for tensor perturbations differs from that of GR.



\section{Concluding remarks} 
\label{sec-discuss}

General Relativity is a successful theory of gravity that describes the evolution of
the universe almost in a satisfactory way. The theory of inflation was found to be an essential addition in this context. Nonetheless, the initial singularity issue, which had been found to be inevitable, inspired to search for other alternatives for inflation where the singularity does not appear. The \textit{Loop Quantum Cosmology} applied to the matter or matter-ekpyrotic bouncing scenario
is an effect of that, which is considered to be a viable alternative to the inflationary paradigm. The model also provides with a bounce of the universe in its early phase and, hence, the singularity problem is naturally avoided. In the present work, considering that the underlying geometry of the universe is best described by the usual spatially flat Friedmann-Lema\^{i}tre-Robertson-Walker line-element, we have shown that a modified version of the mimetic gravity (see \cite{Chamseddine:2013kea, Golovnev:2013jxa} for the introduction of mimetic gravity theory) could be equivalent at the background level to LQC, but however has some important differences at the level of perturbations with LQC in the deformed algebra approach. This probably comes from the fact that it can not come from a covariant Lagrangian where the hypersurface deformation algebra is the classical one, precisely due to the 
deformation of the Dirac's algebra of constraints, which contrast with the fully covariance of the theory presented in this paper. 


\ 
 
The introduction of mimetic gravity in the literature of modern cosmology is very new \cite{Chamseddine:2013kea, Golovnev:2013jxa}, at least in comparison to other cosmological theories and within a few years it has gained a considerable attention in the scientific community  \cite{mukhanov, mukhanov1, Nojiri:2014zqa, Leon:2014yua, Odintsov:2015wwp, Casalino:2018tcd, Brahma:2018dwx}. We have shown that a {  limiting curvature mimetic theory}
through the introduction  of a functional $f(\chi)$ into the Lagrangian, where $\chi$ is the d'Alembertian of the mimetic field, could exhibit some interesting properties based on the choice of $f (\chi)$. At the background level, the { limiting curvature mimetic theory}
may lead to an equivalent structure to that of the LQC for the appropriate  $f$-function. 
While on the 
other hand, concerning the study of the cosmological perturbations {in the flat FLRW spacetime},   the model 
returns different characteristics respectively for the scalar and tensor perturbations. 
For the scalar perturbations, the modified mimetic gravity returns some complicated equation which reflect important differences with respect to LQC, whereas for the tensor perturbations we  
show that the mimetic field exhibits no extra feature in comparison to the General Relativity, that means, at the level of tensor perturbations mimetic gravity coincides with General Relativity.  

\

{
Finally,  it is worthy to point out the results obtained in \cite{bmm} in the framework of Bianchi spacetimes, where the authors show the impossibility of depicting the limiting curvature mimetic gravity as an effective LQC theory, which coincides with our main point in this article.

}

\section*{Acknowledgments}
We would like to thank  {{}  Norbert Bodendorfer, Martin Bojowald and}  Yi-Fu Cai  for useful and stimulating conversations.
This investigation has been supported in part by MINECO (Spain), project MTM2017-84214-C2-1-P, and by the Catalan Government 2017-SGR-247.

\end{document}